\documentclass[runningheads]{llncs}
\usepackage[T1]{fontenc}
\usepackage{epsfig}
\usepackage{graphicx}
\usepackage{amsmath}
\usepackage{amssymb}
\usepackage{multirow}
\usepackage{dblfloatfix}
\usepackage{flushend} 
\usepackage{caption}
\usepackage{hyperref}
\usepackage{indentfirst}

\begin{document}

\title{Deep Learning  based Novel Cascaded Approach for Skin Lesion Analysis}


 \author{Shubham Innani \and
 Prasad Dutande  \and
 Bhakti Baheti  \and
 Ujjwal Baid  \and Sanjay Talbar}
 \authorrunning{S. Innani et al.}
 \institute{Center of Excellence, Signal and Image Prcoessing, Shri Guru Gobind Singhji Institute of Engineering and Technology, Nanded, India \thanks{Authors are greatful to Center of Excellence, Signal and Image Processing , SGGS IET, Nanded, for computing resources} \\
 \url{http://www.sggs.ac.in} \\
 \email{\{2016bec035,prasad.dutande, bahetibhakti, baidujjwal, sntalbar \}@sggs.ac.in}}

\maketitle

\begin{abstract}
    Patients diagnosed with skin cancer like melanoma are prone to a high mortality rate. Automatic lesion analysis is critical in skin cancer diagnosis and ensures effective treatment. The computer-aided diagnosis of such skin cancer in dermoscopic images can significantly reduce the clinicians’ workload and help improve diagnostic accuracy. Although researchers are working extensively to address this problem, early detection and accurate identification of skin lesions remain challenging. This research focuses on a two-step framework for skin lesion segmentation followed by classification for lesion analysis. We explored the effectiveness of deep convolutional neural network (CNN) based architectures by designing an encoder-decoder architecture for skin lesion segmentation and CNN based classification network. The proposed approaches are evaluated quantitatively in terms of the Accuracy, mean Intersection over Union(mIoU) and Dice Similarity Coefficient. Our cascaded end-to-end deep learning-based approach is the first of its kind, where the classification accuracy of the lesion is significantly improved because of prior segmentation.
The code is available at \url{https://www.github.com/shubhaminnani/skin/lesion}


\end{abstract}

\section{Introduction and Related Work}
\begin{figure*}[t]
	\begin{center}
		\includegraphics[scale=0.4]{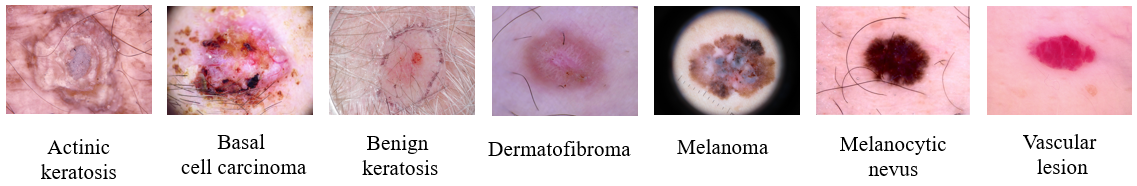}
		\caption{Classwise sample images from HAM10000 classification dataset}
		\label{database_img}
	\end{center}
\end{figure*}

Skin cancer is one of the fatal illnesses in today's world. Even though it is the least common, the disease is responsible for around 91,000 deaths every year until now \cite{wiki2}. Regular monitoring and early detection play a vital function in reducing the mortality rate of skin cancer and can help in precise treatment planning and improving life. Survival rates decrease significantly if skin cancer is left to be treated in an advanced stage of the disease \cite{Rigel}. A dermatologist examines skin images with Dermoscopy, a non-invasive diagnostic tool. This enables dermatologists to visualize delicate clinical designs of skin lesions and subsurface skin structures that are generally not visible to the unaided eye. These images of the skin are studied under a microscope to point out skin abnormalities and classify them into various types of skin cancers \cite{wiki3}. The enhanced dermoscopic images are free from any skin surface reflections, which helps the dermatologist to diagnose skin cancer accurately.

Skin Cancer Detection, i.e., lesion segmentation, is one of the essential and primary steps in accurate and precise treatment planning for various skin diseases. Automatic skin cancer lesion segmentation is very challenging because of the significant variations in lesion characteristics, such as size, location,  shape, color, texture, and skin. The segmentation becomes more arduous due to fuzzy boundaries of lesions, poor color contrast with the surrounding skin, huge intra-class variation, and the existence of antiques such as veins and hairs. Various types of skin cancer images are shown in Fig. \ref{database_img}. Skin lesion segmentation has drawn researchers for over a decade because of its increased clinical applicability and demanding nature. Several image processing and supervised machine learning-based approaches are presented for accurate lesion segmentation, with pros and cons. Most studies pointed toward creating computer-aided design frameworks for skin lesions that would recognize anomalies or skin illnesses. The methods in literature usually follow a certain analysis pipeline \cite{NedaZamani}: The first step is to delineate the lesion area in the image from healthy skin, followed by automated feature extraction to compute the region of interest. The final step is to predict the type of skin lesion (classification task). Several conventional methods are available in the literature to handle the segmentation of skin lesions. The comprehensive review of different lesion segmentation algorithms is available at \cite{CelebiWHIS13} \cite{Celebi_2008} \cite{Tajeddin} \cite{Jahanifar}. \cite{OLIVEIRA2016127}.

In recent times, deep learning techniques have outperformed all the existing state-of-the-art approaches in various computer vision studies like segmentation \cite{ujjwal1} \cite{agri} , detection \cite{bhakti}, classification \cite{ujjwal2}, etc. \cite{prl}\cite{effunet}. The availability of computing resources and huge annotated data has enabled researchers to develop supervised Deep Neural Network models to address these tasks. With the evolution of DCNN and various challenges in skin lesion latterly \cite{ISIC_ISBI_2017}, multiple effective computational approaches have appeared to solve particular problems in this field. Regardless, the most current thriving strategies are based on  CNN \cite{MelanomaRecognition}, \cite{DermaKNet}, \cite{YadingYuan}, \cite{Esteva}. Along with lesion segmentation, deep learning approaches have improved classification performance, leading to better diagnosis of diseases in medical imaging. The methodologies are used to anticipate the existence of illness and recognize the classes. Recent studies demonstrated remarkable performance in classifying skin cancer using deep learning algorithms in binary classification \cite{Esteva} but failed to achieve comparable performance in multi-class classification. 

\begin{figure*}[!t]
	\begin{center}
		\captionsetup{justification=centering}
		\includegraphics[scale=0.3]{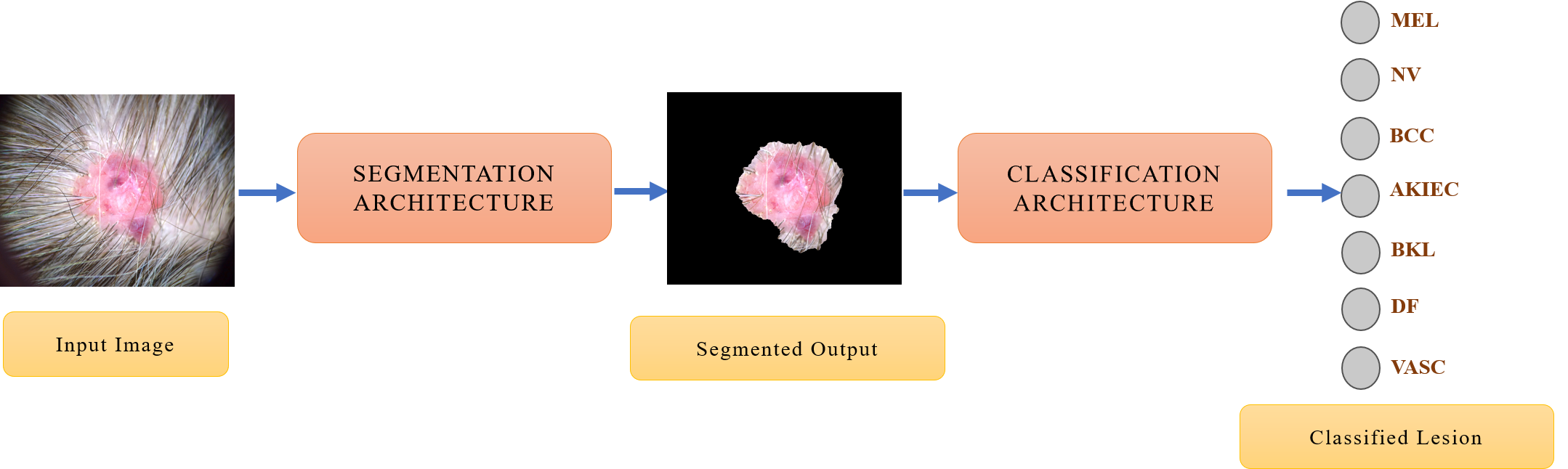}
		\caption{Proposed two stage deep learning based framework for lesion segmentation and classification}
		\label{framework}
	\end{center}
\end{figure*}

This research aims to introduce a two-step automated system that will segment the skin lesion and then classify the disease. After a thorough literature survey, the proposed approach is an end-to-end deep learning-based approach, the first for skin lesion segmentation and classification for seven types of lesions. There is no adequate dataset available having segmentation masks and classification labels in a single dataset for seven different types of lesions. To address this, for segmentation tasks, we work with the International Skin Imaging Collaboration (ISIC) 2018 dataset \cite{ISIC2018} where images with segmentation labels are available, and for classification HAM10000 dataset \cite{Tschandl2018} which consists of seven different skin lesion classes. In our two-step proposed approach, the segmentation task is initially conducted with the ISIC 2018 dataset. With a trained segmentation model, the HAM10000 dataset is segmented, where only classification labels are available. The Region of Interest(ROI) is extracted from segmented images of the HAM10000 dataset fed as input to the classification framework. The two-step framework is shown in Fig. \ref{framework}.

The rest of the article is arranged as follows: Database description is given in Section 2. Presented methods for segmentation and classification of lesions are described in section 3. Section 4 comprises evaluation metrics, experimental results, and performance analysis. This article is concluded in Section 5.

\section{Dataset} \label{dataset}
International Skin Imaging Collaboration (ISIC) 2018 has 2594 images with corresponding ground truth labels for lesion Segmentation. The images have different sizes, from hundred to thousand, and varying width and height ratios. The image lesion has distinct appearances and is located in a different part of the skin. The HAM10000 \cite{Tschandl2018} dataset is used for the classification task, consisting of seven types of lesion disease in the dermoscopy images. Fig. \ref{database_img} provided few images from the dataset. The standard pre-processing like scaling the values between [0,1] or [-1,1] is being implemented on entire dataset.

The classification dataset consists of around 10015 lesions with Actinic keratosis / Bowen’s disease (intraepithelial carcinoma) (AKIEC), Basal cell carcinoma (BCC), Benign keratosis (solar lentigo / seborrheic keratosis/lichen planus-like keratosis) (BKL),  Dermatofibroma (DF), Melanoma (MEL), Melanocytic nevus (NV), Vascular lesion (VASC) diseases. The data distribution is presented in Table \ref{HAM10000}, and we observe that high-class imbalance is challenging in the given datasets, where it is highly skewed towards certain classes. As a result, we observed sparse images for specific groups like DF, VASC, and AKIEC.

\begin{table}[t]
	\captionsetup{justification=centering}
	\caption{Class distribution in HAM10000 dataset}
	\label{HAM10000}
	\centering
	\begin{tabular}{ccc}
		\hline \hline
		\textbf{Class} & \textbf{Number of Images} & \textbf{Class Percentage} \\ \hline \hline 
		AKIEC          & 327                       & 3.27                      \\
		BCC            & 514                       & 5.13                      \\
		BKL            & 1099                      & 10.97                     \\
		DF             & 115                       & 1.15                      \\
		MEL            & 1113                      & 11.11                     \\
		NV             & 6705                      & 66.95                     \\
		VASC           & 142                       & 1.42                      \\ \hline
	\end{tabular}	
\end{table}

\section{Proposed Methodology}
 \begin{figure*}[!b]
	\begin{center}
		\captionsetup{justification=centering}
		\includegraphics[scale=0.38]{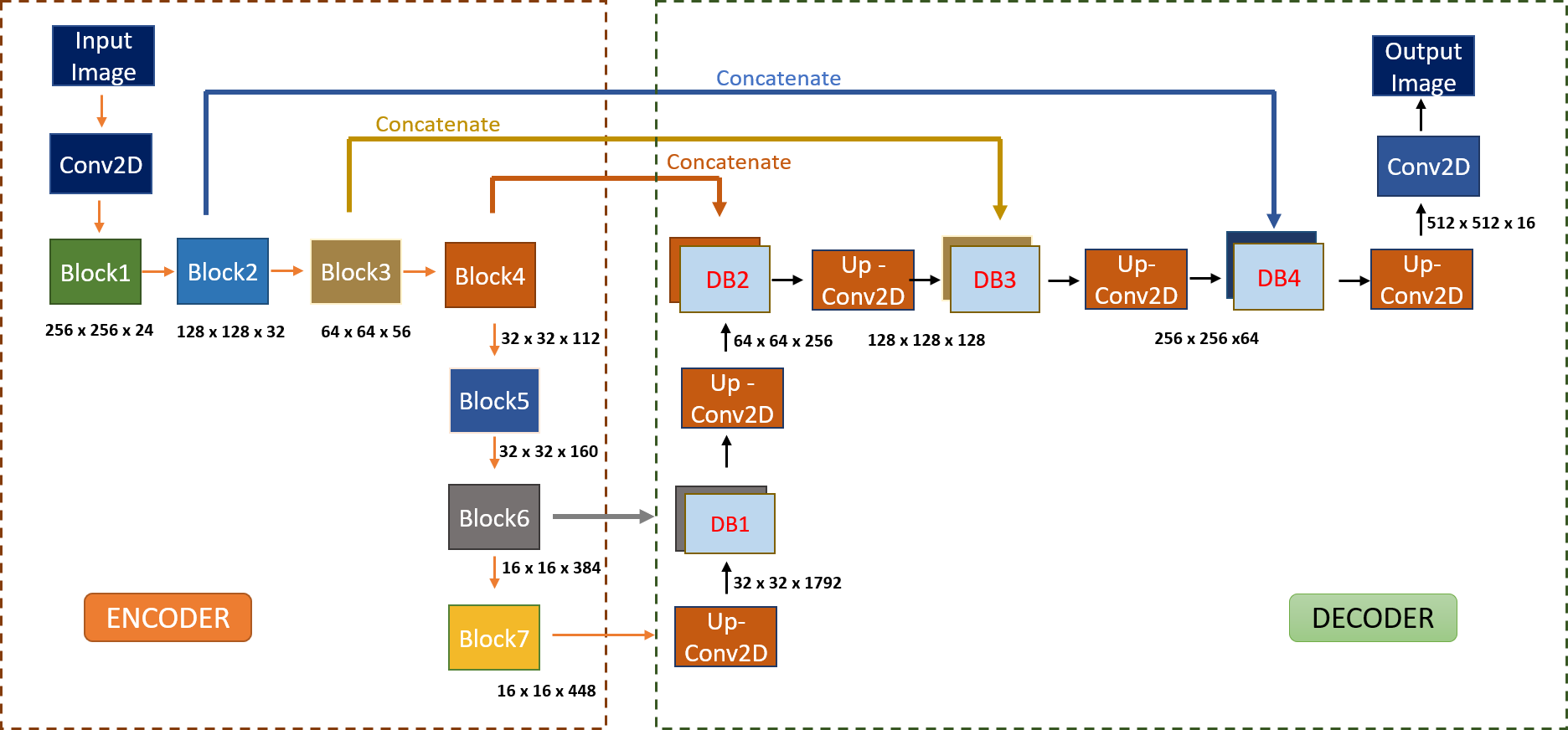}
		\caption{Proposed encoder-decoder architecture for skin lesion segmenation}
		\label{seg}
	\end{center}
\end{figure*}

We propose a two-step framework to handle the task of segmentation and classification in skin lesions. In the first step, the images with skin lesions are segmented to generate coarse-level masks. These segmented masks are multiplied with the corresponding image to extract the coarse level lesion part in the original image, as shown in Fig. \ref{framework}, which removes redundant data in the image, and these ROI images are input to the classification network that signifies the type of lesion.
 
\begin{figure}[!t]
	\begin{center}
		\captionsetup{justification=centering}
		\includegraphics[scale=0.25]{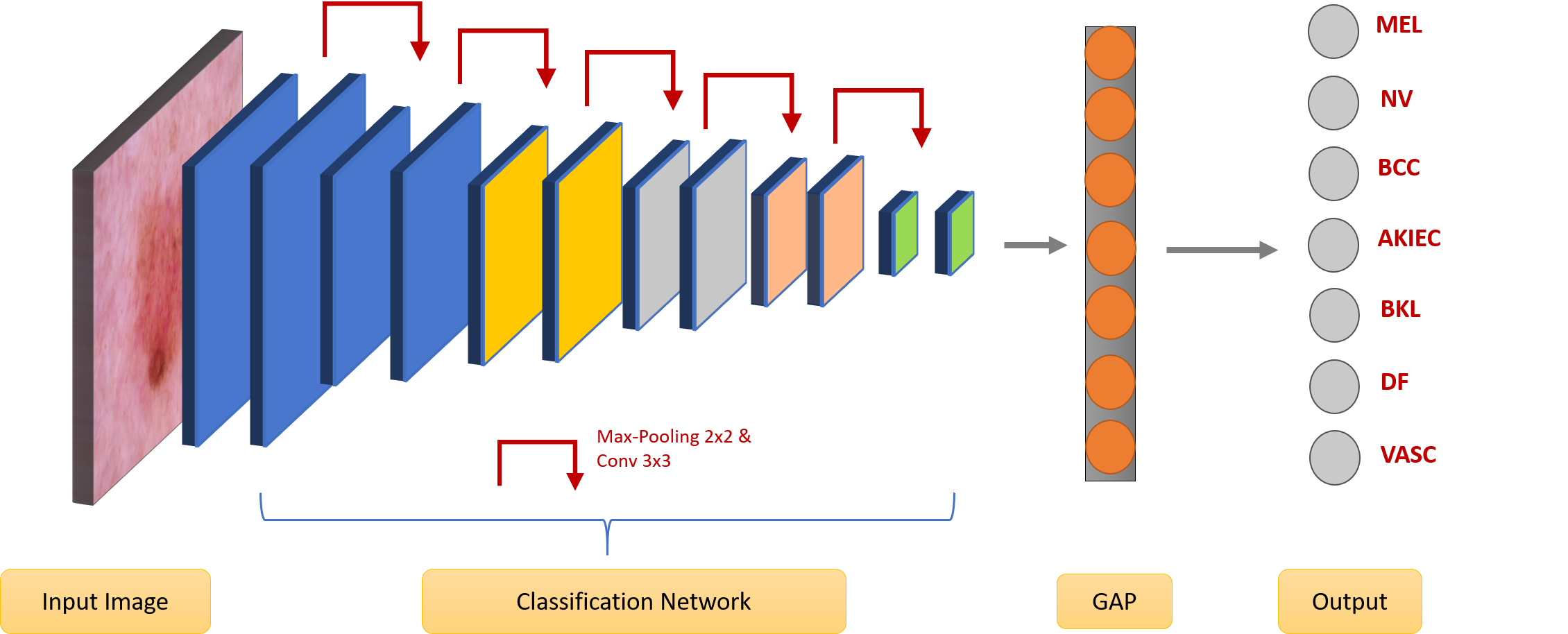}
		\caption{General architecture for skin cancer classifier where Global Average Pooling is abbreviated as GAP.}
		\label{classification}
	\end{center}
\end{figure}

\subsection{Segmentation Approach}

Encoder-decoder architectures are widely used in computer vision for image segmentation task \cite{fusepn} \cite{sultana2020evolution}.   Ronneberger et al. \cite{Ronneberger} presented U-Net, a breakthrough study for medical image segmentation comprising CNN. Generally, a feature learning block is the encoder module to capture spatial features of the input. It downsamples the input image progressively and decreases feature dimensions to catch high-level patterns of the input image. A decoder block consists of layers that upsample the feature map obtained from the encoder output with extracted spatial features. This article's encoder-decoder module is graphically presented in Fig. \ref{seg}. In our approach, we designed three different encoder-decoder networks by replacing the encoder block in U-Net with popular CNN architectures such as ResNet \cite{Kaiming}, InceptionResNetV2 \cite{ResidualInception} and EfficientNets \cite{EfficientNet}. Our architecture based on an encoder-decoder module consists of contraction and expansion paths. The encoder consists of convolutional and max-pooling blocks, which downsample the image to extract high-level features. This CNN output contains denser and high-level feature maps. After every block, the number of feature maps doubles to learn the complex features accurately. 

\begin{figure}[!t]
	\begin{center}
		\captionsetup{justification=centering}
		\includegraphics[scale=0.33]{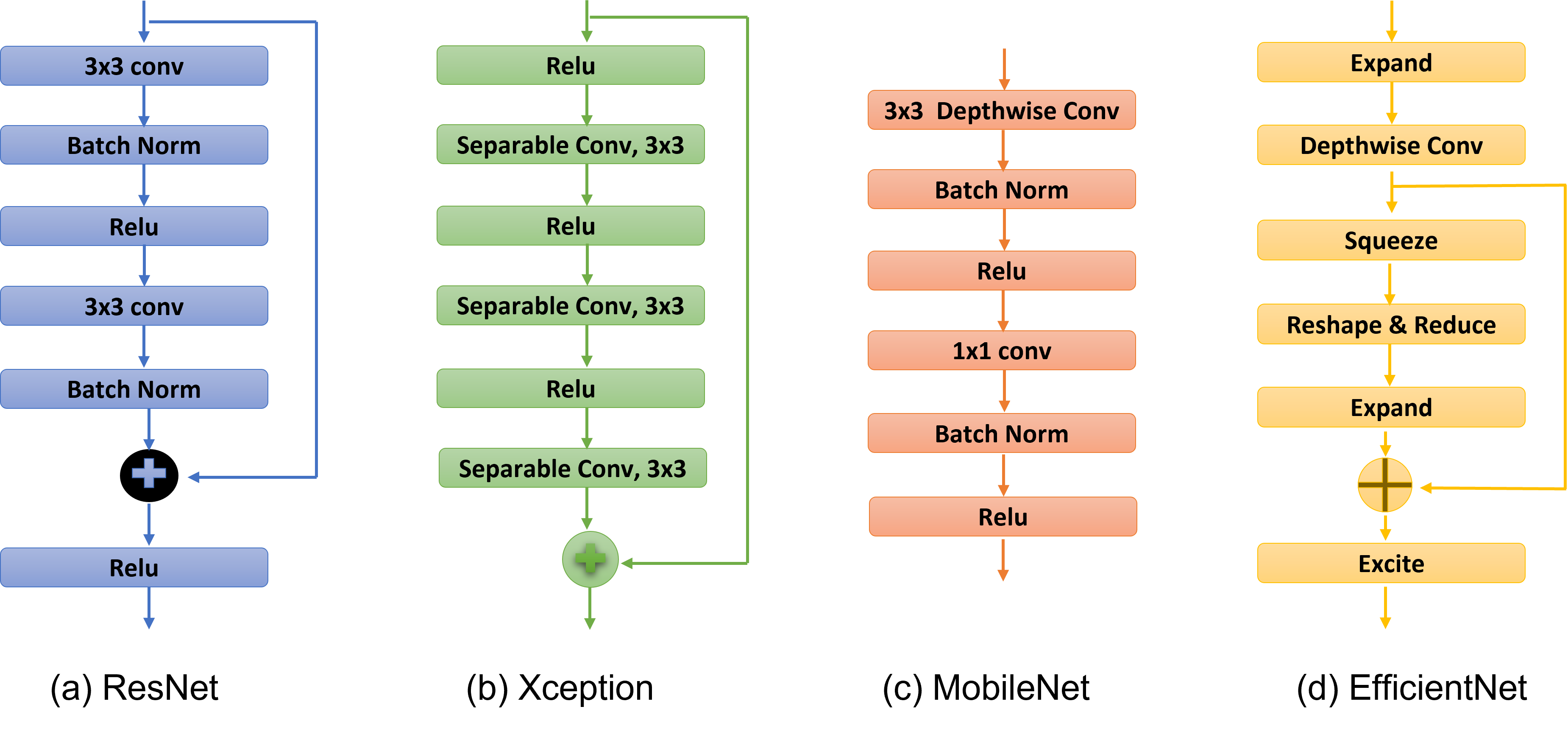}
		\caption{Basic building blocks for various CNN architecture used for classification of skin lesion classes.}
		\label{four}
	\end{center}
\end{figure}
In the encoder, dense features are extracted with an output stride of 16 in each variation, where the output stride is the ratio of the input image shape to the output image shape. These extracted features work well in the classification task, but performance hampers while rebuilding the fine segmentation map. Hence, it is challenging to rebuild the segmentation map of the original input image dimensions from the feature map of the encoder. The decoder module builds the same U-Net decoder architecture to overcome this problem. This encoder output expands in the decoder consisting of convolutional and bilinear upsampling blocks. By concatenating low-level features from the encoder, low-level feature maps are enhanced to the corresponding block of respective size in the decoder to generate the segmented output more precisely.

\subsection{Classification Approach}
Convolutional Neural Network has shown tremendous progress in the task of image classification. With advancements in computational resources and fine-tuning methods, CNN fulfills the demand for performance in terms of accuracy. As shown in Fig. \ref{classification}, a conventional CNN architecture consists of combination blocks of convolutional layer and downsampling layers, followed by a fully connected layer (FC) and the output class. For accurate predictions, CNN automatically pulls the patterns known as features from the input image and carries information at the output block. In the classification step of the dermoscopy, images of the HAM10000 dataset having seven classes from the are to be predicted \cite{Tschandl2018}. We propose to use various classification architectures, which are also used in segmentation encoders like ResNet, Xception, MobileNets, and EfficientNets for classification tasks with an output stride of 32.

\noindent \textbf{ResNets} \cite{Kaiming}: Deep neural network have shattered performance due to the problem of vanishing gradient. To overcome this problem, He et al. proposed the idea of skip connections or residual networks, as shown in Fig. \ref{four}(a). This residual network, known as ResNets, achieved improved performance. ResNet has different variants formed by increasing the residual blocks, namely ResNet18, ResNet50, and so on. ResNet consists of $3\times 3$ convolutional layers stacked with residual or skip connection to form the residual block. For denser prediction and deeper model, maps are periodically doubled. The output of the final layer is 32 times smaller than the input shape. For an image with input shape $224\times224$, the output is of shape $7\times7$. 
 
\noindent \textbf{Xception} \cite{inproceedings}: F. Chollet et al. presented the Xception network as having superior performance. This architecture is inspired by the Inception \cite{ResidualInception}. In Xception, the Inception module in the Inception network is replaced by Depthwise separable convolution (DSC). Xcpetion architecture consisting of 36 convolutional layers grouped in 14 blocks extracts features. All the blocks except the first and last block have skip connections from the previous block. Xception has DSC with residual or skip connection as the primary building layer, as in Fig. \ref{four}(d). The output stride of the final layer is 32.

\noindent \textbf{MobileNet} \cite{howard2017mobilenets}: Mobilenet architecture is a lightweight architecture having depthwise separable convolution as the core layer in building this network, as shown in Fig. \ref{four}(c). DSC is a factorized convolution consisting of pointwise $1\times1$ and depthwise convolution. In mobilenet to each input channel,  a single filter is used depthwise followed by pointwise convolution, fed with input from depthwise convolution for stacking them together. A convolution process combines and filters the input into output in a single step. The DSC is a two-step process in which filtering is carried out in separate layers and combing in another. This division has a significant effect on model size and reduces computation.

\noindent \textbf{EfficientNet} \cite{EfficientNet}: CNNs are developed depending on the resource, and scaling occurs for influential performance while increasing the resources. e.g., ResNet-18 \cite{Kaiming} can be scaled to ResNet-101 by adding some layers. The traditional procedure for scaling the network is to increase the CNN depth or depth or feed with a higher input image size. These methods have proven to improve performance but with tedious manual tuning. In \cite{EfficientNet}, the author proposed a novel approach to scaling the model that uses a compound coefficient that is highly effective for structural scaling of the CNNs. Rather than arbitrarily increasing network dimensions such as resolution, depth, and width, EfficientNet scales every parameter in the compound coefficient with a fixed set of scaling factors. This network is built with mobile inverted bottleneck convolution (MBConv) \cite{MobileNetV2_1} and squeeze and excitation optimization \cite{Squeeze} as shown in Fig. \ref{four}(b).

\section{Result and Discussion}
We randomly divided the ISIC training dataset into 80\% training cohort and 20\% testing cohort. The dataset comprises images of varying sizes rescaled to $512\times512\times3$ for the segmentation task. The segmentation network is trained with a batch size of 8 with a loss function as the sum of cross-entropy and dice loss for 15 epochs setting the parameters for early stopping on Loss. The learning rate was maintained at 0.001 with the ADAM optimizer. In the classification task, we fed the network with an input size of $224 \times 224$ and loss function as categorical cross-entropy. The model is trained with a batch size of 8 for 30 epochs setting the parameters for early stopping on Loss. During training, we initialized the learning rate to 0.001 with the ADAM \cite{adam} optimizer. We augmented the data with various popular augmentation techniques like rotation, shearing, zooming, brightness, and flipping the original images for segmentation and classification tasks. The frameworks are designed with Tensorflow 2.0 and Keras open-source libraries, and the models are trained on NVIDIA P100 GPU with 16 GB memory.

\begin{figure}[!t]
	\begin{center}
		\captionsetup{justification=centering}
		\includegraphics[scale=0.25]{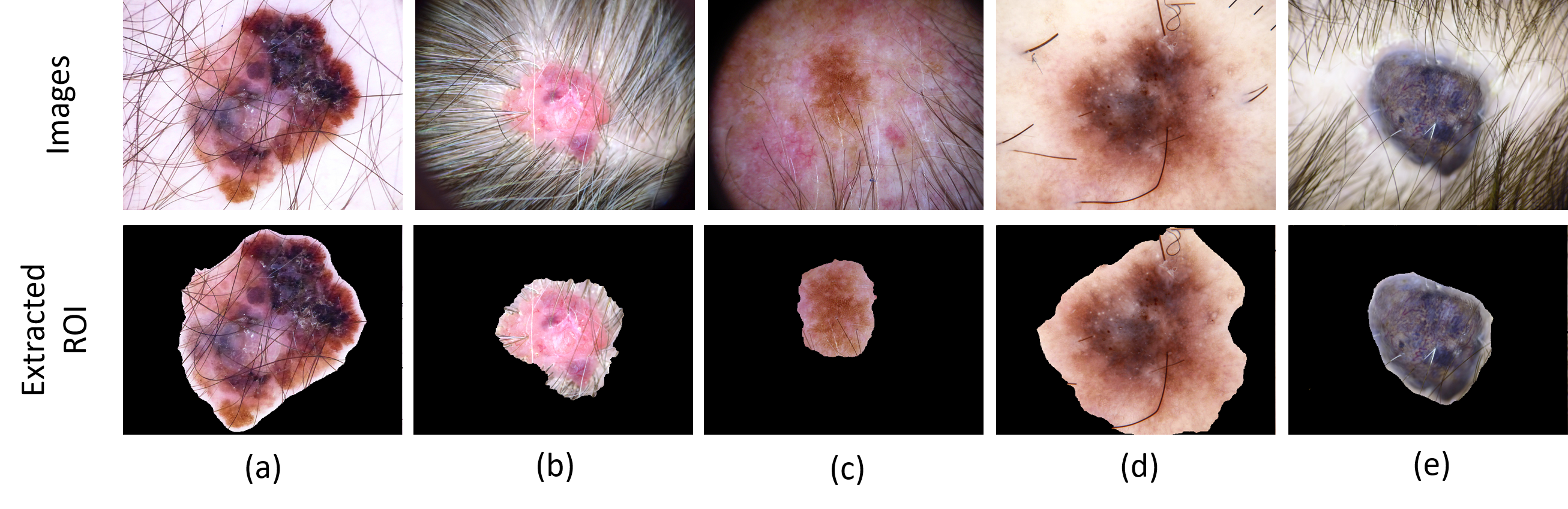}
		\caption{Sample results ROI extracted after segmentation for classification task.}
		\label{seg_op2}
	\end{center}
\end{figure}

\begin{table}[t]
	\centering
	\captionsetup{justification=centering}
	\caption{Performance evaluation of segmentation task on test dataset in terms of Dice Score and Mean Intersection over Union}
	\begin{tabular}{ccc}
		\hline \hline
		Encoder backbone      & Dice Score & mIoU  \\ \hline \hline
		Original U-Net             & 71.53      & 60.58 \\
		ResNet50    & 84.46      & 73.76 \\
		ResNet101   & 86.30      & 76.77 \\
		MobileNet         & 83.90      & 71.32 \\
		InceptionResNetV2 & 87.20      & 78.03 \\
		EfficientNetB4    & 89.56      & 81.42 \\ \hline \hline
	\end{tabular}
	\label{seg_eval_table}
\end{table}

For the segmentation task, U-Net is just a stack of convolutional layers, the original U-Net underperforms in this task. To experiment, we increase the depth of the network with various encoders like ResNet, MobileNet, and EffcientNet. Also, we design an asymmetric decoder, as seen in Fig. \ref{seg}. Concatenation of low-level features occurs at some intervals rather than joining each block from the encoder, as proposed by Ronneberger et al. in U-Net improves performance. The modification improves performance with the proposed deep encoder with the asymmetric decoder.

\begin{table}[t]
	\centering
	\captionsetup{justification=centering}
	\caption{Performance evaluation of classification task on test dataset with and without considering ROI images}
	\begin{tabular}{lcc}
		\hline \hline
		\multirow{2}{*}{\begin{tabular}[c]{@{}l@{}}Classification\\ Architecture\end{tabular}} & \multicolumn{2}{c}{Accuracy}    \\ \cline{2-3} 
		& without ROI    & with ROI       \\ \hline \hline
		ResNet50                                                                               & 74.15          & 78.32          \\
		ResNet101                                                                              & 75.65          & 79.26         \\
		\textbf{MobileNet}                                                                     & 78.54 		   &
		 81.54 			\\
		\textbf{Xception}                                                                      & 78.04 			&
		 82.41 			\\
		EfficientNetB0                                                                         & 75.05          & 79.14          \\
		EfficientNetB3                                                                         & 76.65          & 82.19          \\ \hline \hline
	\end{tabular}
	\label{class_acc}
\end{table}

\begin{figure*}[t]
	\begin{center}
		\captionsetup{justification=centering}
		\includegraphics[scale=0.28]{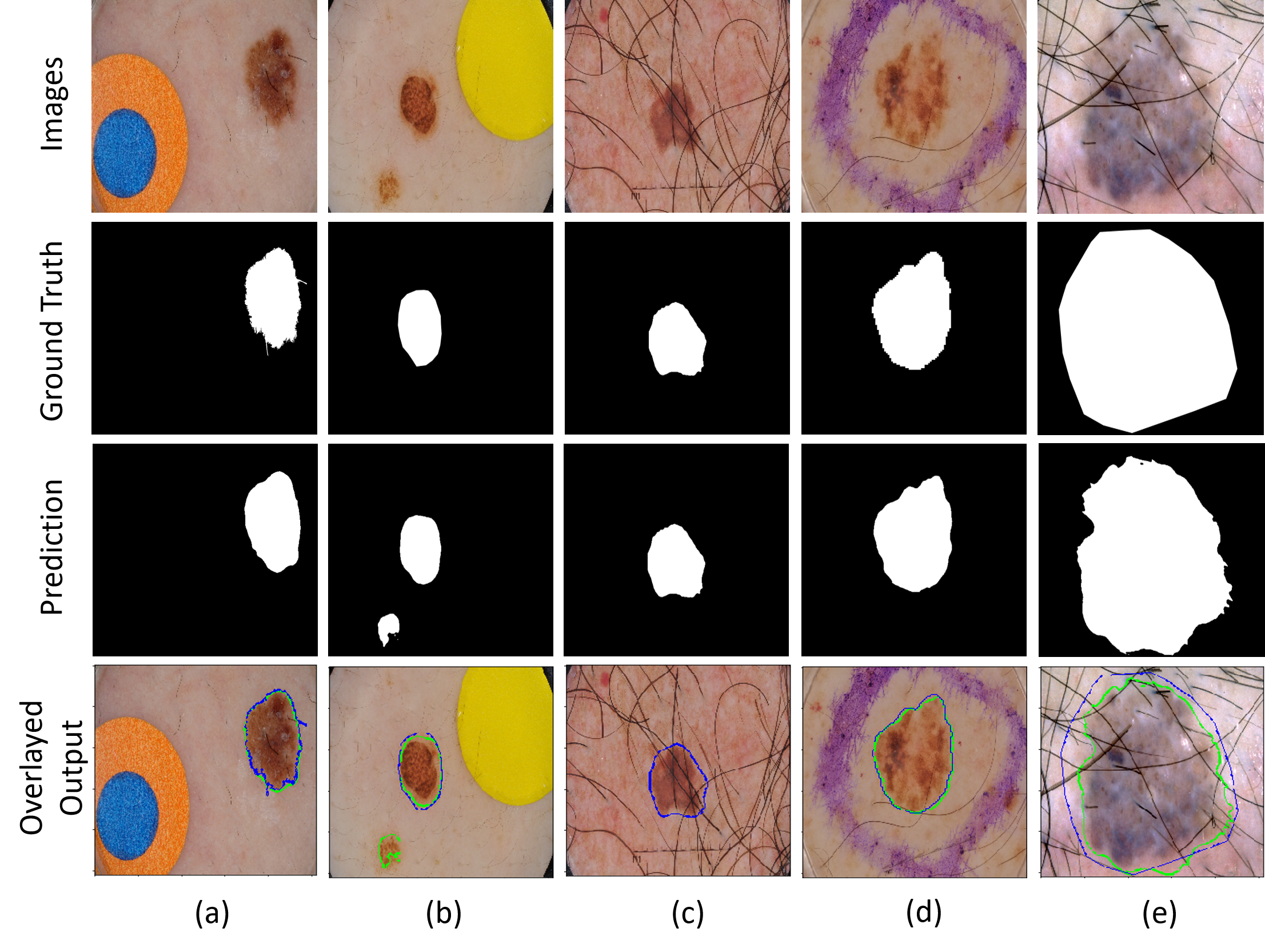}
		\caption{Segmentation Output with EfficientNetB7 as encoder. Each column represents a lesion image, Ground Truth, segmentation predicted by the network and overlayed image of segmented output(Green) and Ground Truth(Blue).}
		\label{seg_op1}
	\end{center}
\end{figure*}

After extracting the ROI by segmenting using an EfficientNet-based encoder, it is fed as input to the various state-of-the-art networks in classification like ResNet, MobileNet, EfficientNet, and Xception. As seen in Table \ref{class_acc}, there is significant performance gain when ROI extracted skin lesion is used. For an in-depth comparison, classification is performed with different CNNs with and without ROI obtained from segmentation. The efficacy of the proposed approaches is evaluated in terms of various popular quantitative evaluation parameters. The performance of segmentation approaches is assessed in terms of the Dice Similarity Coefficient (DSC) and Mean Intersection over Union (mIoU) and the classification approach with accuracy. The performance for the segmentation task of various encoder backbones in DSC and mIoU is given in Table \ref{seg_eval_table}. It can be observed that EfficientNetB4 outperformed other encoders quantitatively. Segmentation outputs predicted by the model for five different images are presented in Fig. \ref{seg_op1}. 

From Fig. \ref{seg_op1}(a) and \ref{seg_op1}(b), it can be observed that the proposed approach performed well even if non-skin objects are present in the image. The architecture could segment lesions, even severe occlusion, because of hairs. These segmentation results are then multiplied with the original image to extract the skin lesion, as shown in Fig. \ref{seg_op2}. It can be observed that besides skin lesions, various surrounding patterns may hamper the classifier learning. ROI from Fig. \ref{seg_op2} (b) and (e) clearly justifies the need of lesion segmentation before classification. The performance evaluation for the classification task with and without ROI is given in Table \ref{class_acc}. The architectures trained on images containing only lesion ROI performed better in terms of accuracy, as shown in Table \ref{class_acc}.


\section{Conclusion}

Skin lesion segmentation and classification are the primary steps in designing the Computer-Aided Diagnostic (CAD) Tool and are essential for precise treatment planning. This study proposed a two-step approach with two distinct databases for skin lesion segmentation and classification. It was observed that, except for lesions, various surrounding patterns might hamper the classifier's learning. To address this, we proposed a two-step approach where in the first step, skin lesions are segmented, and in the second step, ROIs are extracted, which are given input to the classification architecture. Experimentation results showed that classification accuracy with ROI as input outperformed lesion images with surrounding patterns and was improved by 5\%. We currently have the performance of the proposed approach on the publicly available dataset.

{\small
\bibliographystyle{splncs04}
\bibliography{egpaper_for_review}

\begin{thebibliography}{10}
\providecommand{\url}[1]{\texttt{#1}}
\providecommand{\urlprefix}{URL }
\providecommand{\doi}[1]{https://doi.org/#1}

\bibitem{wiki3}
Dermoscopy and mole scans in perth and regional wa.
  \url{https://myskincentre.com.au/service/dermoscopy/} (2018), [Online
  accessed on 20-February-2020]

\bibitem{wiki2}
Melanoma stats, facts, and figures.
  \url{https://www.aimatmelanoma.org/about-melanoma/melanoma-stats-facts-and-figures.}
  (2018), [Online accessed on 20-February-2020]

\bibitem{bhakti}
Baheti, B., Gajre, S., Talbar, S.: Detection of distracted driver using
  convolutional neural network. In: 2018 IEEE/CVF Conference on Computer Vision
  and Pattern Recognition Workshops (CVPRW). pp. 1145--11456 (2018).
  \doi{10.1109/CVPRW.2018.00150}

\bibitem{effunet}
Baheti, B., Innani, S., Gajre, S., Talbar, S.: Eff-unet: A novel architecture
  for semantic segmentation in unstructured environment. In: 2020 IEEE/CVF
  Conference on Computer Vision and Pattern Recognition Workshops (CVPRW). pp.
  1473--1481 (2020). \doi{10.1109/CVPRW50498.2020.00187}

\bibitem{prl}
Baheti, B., Innani, S., Gajre, S., Talbar, S.: Semantic scene segmentation in
  unstructured environment with modified deeplabv3+. Pattern Recognition
  Letters  \textbf{138},  223--229 (2020).
  \doi{https://doi.org/10.1016/j.patrec.2020.07.029},
  \url{https://www.sciencedirect.com/science/article/pii/S0167865520302750}

\bibitem{ujjwal2}
Baid, U., Rane, S.U., Talbar, S., Gupta, S., Thakur, M.H., Moiyadi, A.,
  Mahajan, A.: Overall survival prediction in glioblastoma with radiomic
  features using machine learning. Frontiers in Computational Neuroscience
  \textbf{14} (2020). \doi{10.3389/fncom.2020.00061},
  \url{https://www.frontiersin.org/articles/10.3389/fncom.2020.00061}

\bibitem{ujjwal1}
Baid, U., Talbar, S., Rane, S., Gupta, S., Thakur, M.H., Moiyadi, A., Sable,
  N., Akolkar, M., Mahajan, A.: A novel approach for fully automatic
  intra-tumor segmentation with 3d u-net architecture for gliomas. Frontiers in
  Computational Neuroscience  \textbf{14} (2020).
  \doi{10.3389/fncom.2020.00010},
  \url{https://www.frontiersin.org/articles/10.3389/fncom.2020.00010}

\bibitem{Celebi_2008}
Celebi, M.E., Kingravi, H., Iyatomi, H., Aslandogan, Y., Stoecker, W., Moss,
  R., Malters, J., Grichnik, J., Marghoob, A., Rabinovitz, H., Menzies, S.:
  Border detection in dermoscopy images using statistical region merging. Skin
  research and technology : official journal of International Society for
  Bioengineering and the Skin (ISBS) [and] International Society for Digital
  Imaging of Skin (ISDIS) [and] International Society for Skin Imaging (ISSI)
  \textbf{14},  347--53 (09 2008). \doi{10.1111/j.1600-0846.2008.00301.x}

\bibitem{CelebiWHIS13}
Celebi, M.E., Wen, Q., Hwang, S., Iyatomi, H., Schaefer, G.: Lesion border
  detection in dermoscopy images using ensembles of thresholding methods. CoRR
  \textbf{abs/1312.7345} (2013), \url{http://arxiv.org/abs/1312.7345}

\bibitem{inproceedings}
Chollet, F.: Xception: Deep learning with depthwise separable convolutions. pp.
  1800--1807 (07 2017). \doi{10.1109/CVPR.2017.195}

\bibitem{ISIC_ISBI_2017}
Codella, N.C.F., Gutman, D., Celebi, M.E., Helba, B., Marchetti, M.A., Dusza,
  S.W., Kalloo, A., Liopyris, K., Mishra, N.K., Kittler, H., Halpern, A.: Skin
  lesion analysis toward melanoma detection: {A} challenge at the 2017
  international symposium on biomedical imaging (isbi), hosted by the
  international skin imaging collaboration {(ISIC)}. CoRR
  \textbf{abs/1710.05006} (2017), \url{http://arxiv.org/abs/1710.05006}

\bibitem{ISIC2018}
Codella, N.C.F., Rotemberg, V., Tschandl, P., Celebi, M.E., Dusza, S.W.,
  Gutman, D., Helba, B., Kalloo, A., Liopyris, K., Marchetti, M.A., Kittler,
  H., Halpern, A.: Skin lesion analysis toward melanoma detection 2018: {A}
  challenge hosted by the international skin imaging collaboration {(ISIC)}.
  CoRR  \textbf{abs/1902.03368} (2019), \url{http://arxiv.org/abs/1902.03368}

\bibitem{Esteva}
Esteva, A., Kuprel, B., Novoa, R., Ko, J., Swetter, S., Blau, H., Thrun, S.:
  Dermatologist-level classification of skin cancer with deep neural networks.
  Nature  \textbf{542} (01 2017). \doi{10.1038/nature21056}

\bibitem{DermaKNet}
{González-Díaz}, I.: Dermaknet: Incorporating the knowledge of dermatologists
  to convolutional neural networks for skin lesion diagnosis. IEEE Journal of
  Biomedical and Health Informatics  \textbf{23}(2),  547--559 (March 2019).
  \doi{10.1109/JBHI.2018.2806962}

\bibitem{Kaiming}
He, K., Zhang, X., Ren, S., Sun, J.: Deep residual learning for image
  recognition. CoRR  \textbf{abs/1512.03385} (2015),
  \url{http://arxiv.org/abs/1512.03385}

\bibitem{howard2017mobilenets}
Howard, A.G., Zhu, M., Chen, B., Kalenichenko, D., Wang, W., Weyand, T.,
  Andreetto, M., Adam, H.: Mobilenets: Efficient convolutional neural networks
  for mobile vision applications (2017)

\bibitem{Squeeze}
Hu, J., Shen, L., Sun, G.: Squeeze-and-excitation networks. CoRR
  \textbf{abs/1709.01507} (2017), \url{http://arxiv.org/abs/1709.01507}

\bibitem{agri}
Innani, S., Dutande, P., Baheti, B., Talbar, S., Baid, U.: Fuse-pn: A novel
  architecture for anomaly pattern segmentation in aerial agricultural images.
  In: 2021 IEEE/CVF Conference on Computer Vision and Pattern Recognition
  Workshops (CVPRW). pp. 2954--2962 (2021). \doi{10.1109/CVPRW53098.2021.00331}

\bibitem{fusepn}
Innani, S., Dutande, P., Baheti, B., Talbar, S., Baid, U.: Fuse-pn: A novel
  architecture for anomaly pattern segmentation in aerial agricultural images.
  In: 2021 IEEE/CVF Conference on Computer Vision and Pattern Recognition
  Workshops (CVPRW). pp. 2954--2962 (2021). \doi{10.1109/CVPRW53098.2021.00331}

\bibitem{Jahanifar}
{Jahanifar}, M., {Zamani Tajeddin}, N., {Mohammadzadeh Asl}, B., {Gooya}, A.:
  Supervised saliency map driven segmentation of lesions in dermoscopic images.
  IEEE Journal of Biomedical and Health Informatics  \textbf{23}(2),  509--518
  (March 2019). \doi{10.1109/JBHI.2018.2839647}

\bibitem{adam}
Kingma, D., Ba, J.: Adam: A method for stochastic optimization. International
  Conference on Learning Representations  (12 2014)

\bibitem{OLIVEIRA2016127}
Oliveira, R.B., Filho, M.E., Ma, Z., Papa, J.P., Pereira, A.S., Tavares,
  J.M.R.: Computational methods for the image segmentation of pigmented skin
  lesions: A review. Computer Methods and Programs in Biomedicine
  \textbf{131},  127 -- 141 (2016).
  \doi{https://doi.org/10.1016/j.cmpb.2016.03.032},
  \url{http://www.sciencedirect.com/science/article/pii/S0169260716303418}

\bibitem{Rigel}
Rigel, D.S., Russak, J., Friedman, R.: The evolution of melanoma diagnosis: 25
  years beyond the abcds. CA: A Cancer Journal for Clinicians  \textbf{60}(5),
  301--316. \doi{10.3322/caac.20074},
  \url{https://acsjournals.onlinelibrary.wiley.com/doi/abs/10.3322/caac.20074}

\bibitem{Ronneberger}
Ronneberger, O., Fischer, P., Brox, T.: U-net: Convolutional networks for
  biomedical image segmentation. In: Navab, N., Hornegger, J., Wells, W.M.,
  Frangi, A.F. (eds.) Medical Image Computing and Computer-Assisted
  Intervention -- MICCAI 2015. pp. 234--241. Springer International Publishing,
  Cham (2015)

\bibitem{MobileNetV2_1}
Sandler, M., Howard, A.G., Zhu, M., Zhmoginov, A., Chen, L.: Inverted residuals
  and linear bottlenecks: Mobile networks for classification, detection and
  segmentation. CoRR  \textbf{abs/1801.04381} (2018),
  \url{http://arxiv.org/abs/1801.04381}

\bibitem{sultana2020evolution}
Sultana, F., Sufian, A., Dutta, P.: Evolution of image segmentation using deep
  convolutional neural network: A survey. arXiv preprint arXiv:2001.04074
  (2020)

\bibitem{Tajeddin}
{Tajeddin}, N.Z., {Asl}, B.M.: A general algorithm for automatic lesion
  segmentation in dermoscopy images. In: 2016 23rd Iranian Conference on
  Biomedical Engineering and 2016 1st International Iranian Conference on
  Biomedical Engineering (ICBME). pp. 134--139 (Nov 2016).
  \doi{10.1109/ICBME.2016.7890944}

\bibitem{NedaZamani}
Tajeddin, N.Z., Asl, B.M.: Melanoma recognition in dermoscopy images using
  lesion's peripheral region information. Computer Methods and Programs in
  Biomedicine  \textbf{163},  143 -- 153 (2018).
  \doi{https://doi.org/10.1016/j.cmpb.2018.05.005},
  \url{http://www.sciencedirect.com/science/article/pii/S0169260717313251}

\bibitem{EfficientNet}
Tan, M., Le, Q.V.: Efficientnet: Rethinking model scaling for convolutional
  neural networks. CoRR  \textbf{abs/1905.11946} (2019),
  \url{http://arxiv.org/abs/1905.11946}

\bibitem{Tschandl2018}
Tschandl, P., Rosendahl, C., Kittler, H.: {Data descriptor: The HAM10000
  dataset, a large collection of multi-source dermatoscopic images of common
  pigmented skin lesions}. Scientific Data  \textbf{5}, ~1--9 (2018).
  \doi{10.1038/sdata.2018.161}

\bibitem{MelanomaRecognition}
{Yu}, L., {Chen}, H., {Dou}, Q., {Qin}, J., {Heng}, P.: Automated melanoma
  recognition in dermoscopy images via very deep residual networks. IEEE
  Transactions on Medical Imaging  \textbf{36}(4),  994--1004 (April 2017).
  \doi{10.1109/TMI.2016.2642839}

\bibitem{YadingYuan}
Yuan, Y., Lo, Y.: Improving dermoscopic image segmentation with enhanced
  convolutional-deconvolutional networks. CoRR  \textbf{abs/1709.09780} (2017),
  \url{http://arxiv.org/abs/1709.09780}

\bibitem{ResidualInception}
{Zhang}, X., {Huang}, S., {Zhang}, X., {Wang}, W., {Wang}, Q., {Yang}, D.:
  Residual inception: A new module combining modified residual with inception
  to improve network performance. In: 2018 25th IEEE International Conference
  on Image Processing (ICIP). pp. 3039--3043 (Oct 2018).
  \doi{10.1109/ICIP.2018.8451515}

\end{thebibliography}
}

\end{document}